\documentclass[aps,prl,twocolumn,showpacs,preprintnumbers,groupedaddress]{revtex4}
\usepackage{epsfig,graphics,amssymb,amsmath,subeqnarray}

\begin{document}

\title{Hydrodynamic phase-locking of swimming microorganisms}
\author{Gwynn J Elfring}
\author{Eric Lauga}
\email{elauga@ucsd.edu}
\affiliation{
Department of Mechanical and Aerospace Engineering, 
University of California San Diego,
9500 Gilman Drive, La Jolla CA 92093-0411, USA.}
\date{\today}
\begin{abstract}
Some microorganisms, such as spermatozoa,  synchronize their flagella when swimming in close proximity. Using a simplified model (two infinite, parallel, two-dimensional waving sheets), we show that phase-locking arises from hydrodynamics forces alone, and has its origin in the front-back asymmetry of the geometry of their flagellar waveform. The time-evolution of the phase difference between co-swimming cells depends only on the nature of this geometrical asymmetry, and  microorganisms can phase-lock into conformations which minimize or maximize energy dissipation.

\end{abstract}


\maketitle

Swimming cells, such as spermatozoa or flagellated bacteria, are ubiquitous in nature, yet their dynamics  in a world without inertia is often counter-intuitive  \cite{purcell}. Much insight has been gained in the past about the nature of the swimming of flagellated microorganisms, from the pioneering work of Taylor \cite{taylor1}, through to many detailed reviews on locomotion at the microscale \cite{lighthill,brennan,lauga2}. Spermatozoa in particular have received much attention  in an effort to improve our understanding of the biomechanics of reproductive processes \cite{fauci06}.

One particularly puzzling phenomenon observed in swimming spermatozoa and other microorganisms is the apparent synchronization of the beating of the flagella  of two or more cells when in close proximity \cite{taylor1,yang}. 
This phenomenon was first modeled by Taylor using infinite two-dimensional sinusoidal sheets \cite{taylor1}. Taylor found that the most energetically efficient configuration for two swimmers close together was to beat in synchrony.  A computational model of the same setup showed that at small but finite Reynolds numbers, the sheets can achieve stable phase-locking at in-phase and opposite-phase configurations \cite{fauci1}, a result which remains valid for finite swimmers with flagella of linearly increasing amplitude  \cite{fauci2}. Computations in two dimensions showed that the flow fields of interacting swimmers tend to cluster them together into tight synchronized groups  \cite{yang}. Large arrays of cilia (short, closely-packed flagella) also display synchronization if the internal force-generation mechanism generating their beat pattern is load-dependant \cite{gueron1,vilfan}.

What is still not understood is what constitutes the essential physical ingredients to obtain an evolution in time to a phase-locked configuration  between cells swimming close to each other. Here we consider a simplified model of nearby swimming cells with a prescribed waveform. We show that stable phase-locking can be obtained purely passively, due to  hydrodynamic interactions. The phase-locked state to which the cells evolve is dictated solely by the geometry of the flagellar waveforms of the cells (specifically, their front-back asymmetry), and not by considerations of energy dissipation.

In the spirit of Taylor \cite{taylor1}, we consider a model of co-swimming cells consisting of two infinite parallel two-dimensional sheets propagating lateral waves of transverse oscillations with prescribed wavenumber $k$,  frequency $\omega$, and wave speed $c=\omega/k$; each sheet is thereby propelled in the direction opposite to the wave. 
This idealized geometrical model, which has been used successfully in the past to study other properties of cell locomotion \cite{lauga2}, will allow us to clearly elucidate  the necessary ingredients required for synchronization. We also relate it below to experimental observations.

\begin{figure}[t]
\includegraphics[width=0.49\textwidth]{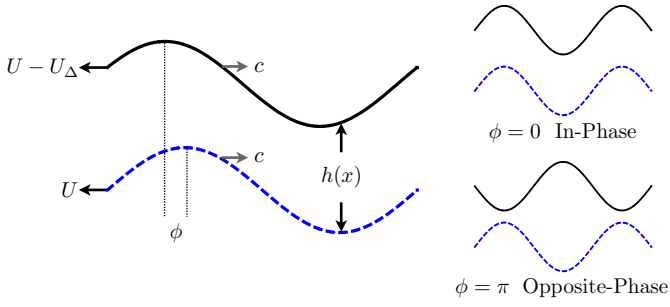}
\caption{(Color online). 
Our model for flagellar phase-locking:  Two infinite, parallel, 2D sheets propagate periodic waves at a speed $c$ leading to swimming in  the opposite direction. The top sheet is allowed to move with velocity $U_\Delta$ with respect to the bottom sheet and is out of phase by an angle $\phi$.}
\label{system}
\end{figure}

The  shape of the waveform is assumed to be the same for both swimmers, and is described  by an arbitrary function $a$. The position of the bottom sheet relative to an axis about which it is centered vertically,  is denoted $y_1=a(kx-\omega t)$ in its swimming frame. The top sheet which is some mean distance $\bar{h}$ above and parallel to the bottom sheet moves at a speed $U_\Delta$ relative to the bottom sheet. The two sheets have an instantaneous phase difference denoted $\phi$  (see Fig.~\ref{system}); $U_\Delta$ is defined to be positive if the upper sheet moves to the right relative to the lower sheet; $\phi$ is defined to be positive if the upper sheet is left of the lower sheet by $\phi$. The instantaneous position of the top sheet is thus given by $y_2 = \bar{h}+a(kx-\omega t - \int_{0}^{t} k U_\Delta(t')dt' + \tilde{\phi})$, where $\tilde{\phi}$ is the phase difference at $t=0$.

The governing hydrodynamics equations for low-Reynolds number flow in an incompressible Newtonian fluid are the Stokes equations, $\{\boldsymbol{\nabla}p=\mu\nabla^2\mathbf{u},\,\boldsymbol \nabla\cdot \mathbf{u}=0\}$, for the velocity field, $\mathbf{u}=(u,v)$ and pressure, $p$. We non-dimensionalize using $\hat{x}^*=xk$, $y^*=y/\bar{h}$, $t^*=t\omega$, $u^*=u/c$, $U_\Delta^*=U_\Delta/c$, $v^*=v/\epsilon c$, $p^*=p\epsilon^2/\mu\omega$, where $\epsilon$ indicates the ratio of mean separation of swimmers to their wavelength, $\epsilon=\bar{h}k$. Fluid force per unit width are nondimensionalized as  $f^* =\epsilon f/\mu c$ and energy dissipation rate per unit width as $\dot{E}^*= \epsilon^2\dot{E}/\mu\omega c\bar{h}$. 
For convenience we introduce the variables $h^*=y_2^*-y_1^*$, $x^*=\hat{x}^*-t^*$, $\phi=\tilde{\phi}-\int_{0}^{t^*} U_\Delta^*(t')dt'$. Consequently, we have $y_1^*=a^*(x^*)$, and $y_2^*=1+a^*(x^*+\phi)$, and  the phase evolves in time according  to $\dot{\phi}=-U_\Delta^*$. We further assume that the waveform possesses reflectional symmetry with respect to the horizontal axis, namely $a^*(x^*+\pi)=-a^*(x^*)$, in order to focus on cells swimming along straight lines \cite{goldstein77}. We now drop the ($^*$) notation for convenience,  and refer to dimensionless variables.

As seen experimentally, phase-locking can occur when the cells flagella beat  close together, therefore an appropriate limit to study is when the mean distance between them is much smaller than their wavelength, i.e. $\epsilon\ll 1$. In this limit the Stokes equations reduce to the lubrication equations, $\partial p/ \partial x = \partial^2 u / \partial y^2$ and $\partial p / \partial y=0$ \cite{batchelor}. We solve these equations in a frame moving with the waveform of the bottom sheet \cite{chan}. The boundary conditions are hence $u(y_1)=-1$ and $u(y_2)=-1+U_\Delta$. 
The solution is for $u$ readily obtained as
\begin{equation}
u = \frac{1}{2}\frac{dp}{dx}(y-y_2)(y-y_1)+U_\Delta\frac{y-y_1}{y_2-y_1}-1 . \label{usol}
\end{equation}
Integrating the continuity equation over $h$ yields a relation between the gradient of the flow rate between the sheets, $Q=\int_{y_1}^{y_2}udy$,  and their relative motion as
\def\d{{\rm d}}
\begin{equation}
\frac{\partial Q}{\partial x}=U_\Delta \frac{\partial y_2}{\partial x}\cdot
 \label{cont}
\end{equation}

In order to determine the physical conditions for phase-locking to occur, we first set $U_\Delta=0$ and investigate the resultant horizontal hydrodynamic force, $f_x$, acting on the upper sheet. In a free-swimming situation, the upper sheet would move at a rate such that the viscous drag would balance with $f_x$ (see below). With $U_\Delta=0$, we know from Eq.~\eqref{cont} that $Q$ is constant and upon integrating Eq.~\eqref{usol} over $h$ we get $Q=-h-(h^3/12)(dp/dx)$. Since the system is $2\pi$ periodic, we have $\int_{0}^{2\pi} (dp/dx) \d x=0$, which leads to  $Q=-I_2/I_3$ where $I_j=\int_{0}^{2\pi}h^{-j}\d x$. The pressure gradient is then obtained to be
\begin{equation}
\frac{dp}{dx}=12\left(\frac{I_2}{I_3h^3}-\frac{1}{h^2}\right)\cdot \label{pressure}
\end{equation}
The force per unit width is determined by integrating the stress over the upper sheet, $f_x={\bf e}_x\cdot \int_S \boldsymbol{\sigma}\cdot\mathbf{n}\ ds$, where $\bf n$ is the unit normal to the sheet into the fluid and $\boldsymbol{\sigma} = -p {\bf 1} + \mu (\nabla {\bf u} + \nabla {\bf u}^T )$ is the stress tensor. 
Using integration by parts, the force is given by
\begin{equation}
f_x  =  \int_0^{2\pi} \left(y_2\frac{dp}{dx}- \frac{\partial u}{\partial y}\right)\mid_{y=y_2}\d x . \label{force_gen}
\end{equation}
Using Eqs.~\eqref{usol} and \eqref{pressure}, we finally obtain the force
\begin{equation}
f_x=6\int_{0}^{2\pi}\left\{\left(\frac{I_2}{I_3h^3}-\frac{1}{h^2}\right)\left[a(x+\phi)+a(x)\right]\right\}\d x. \label{force}
\end{equation}

Physical insight can be gained by inspection of Eq.~\eqref{force}. When $\phi=0$ (in-phase swimming), $h$ is constant, and the force is identically zero for all waveforms $a(x)$. Furthermore, since our waveforms possess reflection symmetries about the horizontal axis, the force is also exactly zero when $\phi=\pi$ (opposite-phase swimming). What is however the nature of the hydrodynamic force about the fixed points $\phi=0,\pi$?

\begin{figure}
\includegraphics[width=0.49\textwidth]{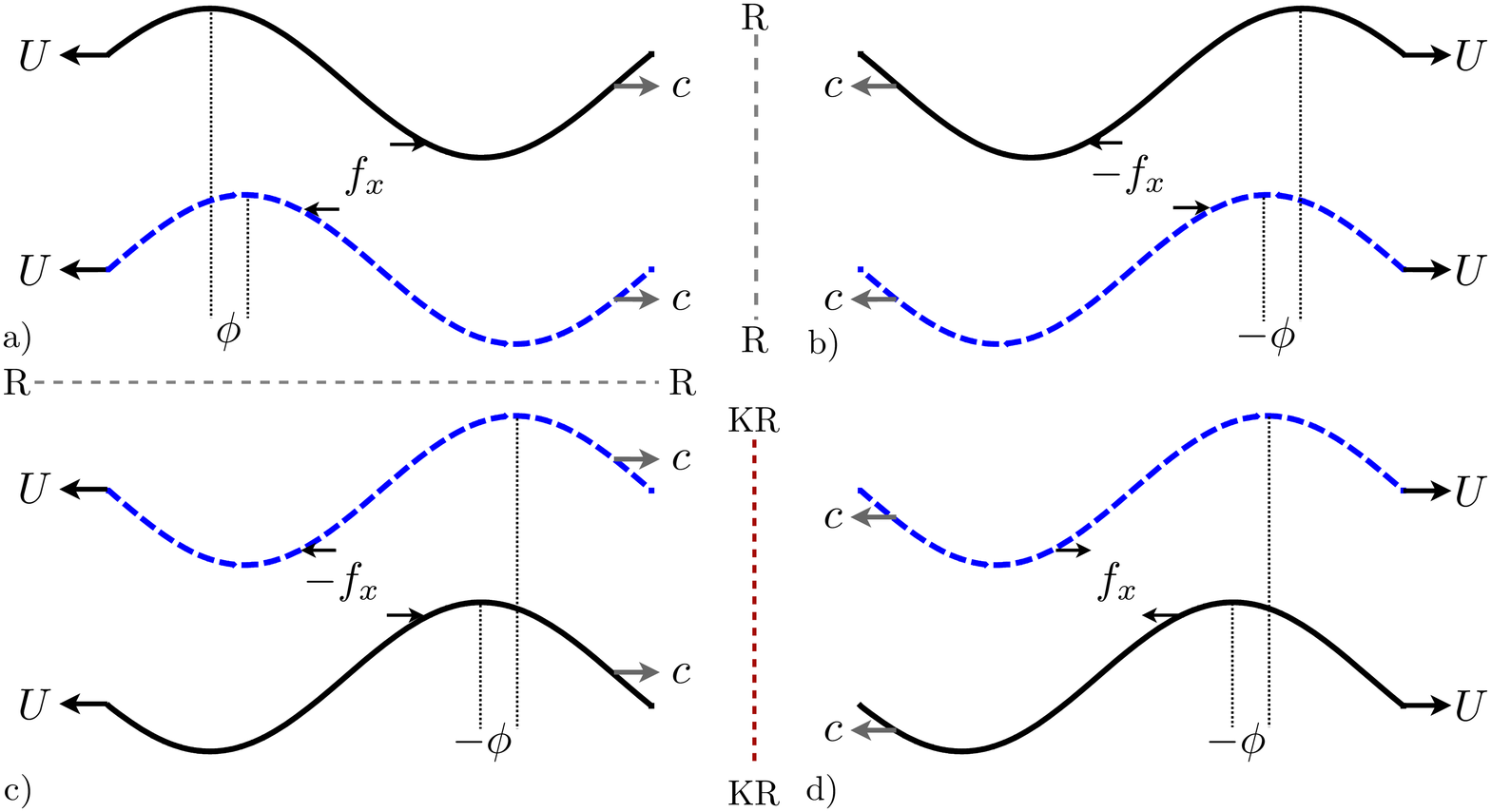}
\caption{(Color online). Swimmers with reflection symmetries by the horizontal and vertical axes cannot phase lock: 
If a  relative force exists  in \ref{sines}a, we obtain the forces in \ref{sines}b and \ref{sines}c by reflection symmetries (R planes). Applying kinematic reversibility  to \ref{sines}c (KR line) leads to a force in \ref{sines}d which is minus the one in \ref{sines}b, indicating that they both must be zero.
}
\label{sines}
\end{figure}

Using symmetry arguments, we can first demonstrate that if the waveforms also possess reflection symmetry with respect to the vertical axis (such as a pure sinewave), no phase locking is ever possible. This is illustrated in Fig.~\ref{sines}. Suppose the force between the swimmers acts to stabilize the phase difference (Fig.~\ref{sines}a). Let us then compare  the forces on the  setup obtained  by a reflection by the vertical axis (Fig.~\ref{sines}b), with that  obtained first by  reflection by the horizontal axis and then by kinematic reversibility (i.e. change of the direction of the wave)  (Fig.~\ref{sines}c and d). The force in Fig.~\ref{sines}d is destabilizing while the one in Fig.~\ref{sines}b is, for the same setup, stabilizing, indicating that both of them must be zero. In particular, sine-waves, such as the ones considered in  Ref.~\cite{taylor1,fauci1}, cannot phase-lock. Note that this argument holds also for finite flagella with a shape invariant upon reflection about the vertical axis versus the horizontal axis, such as sine-waves with an integer number of wavelengths.

If the waveform is not front-back symmetric (i.e. lacks reflection symmetry with respect to the vertical axis), such as spermatozoa  with flagellar waveforms of increasing amplitude \cite{rikmenspoel}, the comparison between Fig.~\ref{sines}b and Fig.~\ref{sines}d cannot be made, and a force can appear. In that case, the comparison between Fig.~\ref{sines}a and Fig.~\ref{sines}c shows that this force must be an odd function of the phase difference, i.e. $f_x(-\phi)=-f_x(\phi)$. For small variations about the fixed points, $\phi=\phi_0+\phi'$ with $\phi'\ll 1$,  the force given by Eq.~\eqref{force} 
 determines the stability of $\phi_0$. Near the in-phase configuration ($\phi_0=0$)  the force is
\begin{equation}
f_{x_0} \approx -72\phi'^3\int_{0}^{2\pi}a\left(\frac{da}{dx}\right)^3 \d x,
\label{force_in}
\end{equation}
to leading order in $\phi'$. If $\delta$ denotes the amplitude of the waveform, we see that $f_{x_0} \sim\pm \phi'^3\delta^4$, with a sign that depends solely on the wave geometry; the sign $+$ ($-$) leads to stability (instability) of in-phase swimming. Similarly, expanding about  opposite-phase swimming ($\phi_0=\pi$), we get at  leading order
\begin{equation}
f_{x_\pi}\approx6 \phi'^3\int_{0}^{2\pi}\frac{(da/dx)^3}{(1-2a)^4}\left(\frac{2}{(1-2a)}\frac{J_2}{J_3}-1\right) \d x,
 \label{force_out}
\end{equation}
where $J_n=\int_{0}^{2\pi}(1-2a)^{-n}\d x$.  For small-amplitude  waves with $\delta \ll 1$,  Eq.~\eqref{force_out} simplifies to 
\begin{equation}
f_{x_\pi} \approx 72\phi'^3 \int_{0}^{2\pi}a\left(\frac{da}{dx}\right)^3 \d x.
\label{force_out2}
\end{equation}
Comparing Eq.~\eqref{force_in} with Eq.~\eqref{force_out2}, we see that  forces near in-phase and opposite-phase configurations have the same magnitude, but opposite signs. 
One of the fixed points for phase-locking is therefore stable while  the other is unstable, in a manner which depends solely on the  waveform geometry: If the waveform $a$ is such that  $A \equiv  \int_{0}^{2\pi}a\left({da}/{dx}\right)^3 \d x < 0$ ($>0$) then  in-phase swimming is stable (unstable) while opposite-phase swimming is unstable (stable)  
\footnote{From  kinematic reversibility we get that a stable fixed point for forward swimming  is unstable for backwards swimming.}.

The rate of energy dissipated in the fluid between the  sheets per unit width is  $\dot{E}=\int\!\!\! \int \boldsymbol{\sigma}:\boldsymbol{\nabla}\mathbf{u}\ \d y\d x$, 
which  is given over one wavelength  by
\begin{equation}
\dot{E}=12\int_{0}^{2\pi}h^3\left(\frac{I_2}{I_3h^3}-\frac{1}{h^2}\right)^2\d x. 
\label{energy}
\end{equation}
For small-amplitude waves near in-phase swimming ($\phi_0=0$), we get $\dot{E}_0\approx12\phi'^2\int_{0}^{2\pi}(da/dx)^2 \d x$,  while near  the opposite-phase configuration ($\phi_0=\pi$), we obtain $\dot{E}_\pi\approx12\int_{0}^{2\pi}\left(4a^2-(da/dx)^2\phi'^2\right)\d x$. In-phase swimming is therefore always the situation where the swimmers have to do the least amount of work, while opposite-phase the most work \cite{taylor1,fauci1} 
 \footnote{Since the forces reverse upon a reversal of swimming direction, the energy dissipated is invariant under such a reversal.}.
Comparing this result with the forces in Eqs.~\eqref{force_in} and \eqref{force_out2}, we see explicitly that there is no relationship between viscous dissipation and hydrodynamic force, and the  swimmers can be forced into a stable conformation where the energy dissipation is in fact maximum (when $A>0$).

In order to observe the evolution of the phase angle towards a phase-locked state, we now allow $U_\Delta$ to be nonzero. Evaluating Eq.~\eqref{cont} with Eq.~\eqref{usol} and integrating in $x$ we get 
\begin{equation}
\frac{dp}{dx}=\frac{6U_\Delta-12}{h^2}-\frac{12U_\Delta y_2+C}{h^3} \label{pressure2},
\end{equation}
where $C$ is a constant,  found by enforcing that the pressure is $2\pi$ periodic, $C=(6U_\Delta-12)(I_2/I_3)-12U_\Delta(K/I_3)$, where $K=\int_{0}^{2\pi}y_2/h^3 \d x$. We then obtain $U_\Delta$ by imposing that the  swimmers are force-free. Substituting Eqs.~\eqref{pressure2} and \eqref{usol} into Eq.~\eqref{force_gen} with $f_x=0$, we  solve for $\dot{\phi} = -U_\Delta$ and get
\begin{equation}
\frac{d\phi}{dt}=-\Lambda f_x^s ,
\label{phi_dot}
\end{equation}
where
\begin{align}
\Lambda^{-1} = \int_0^{2\pi} &\left\{\left[ \frac{3}{h^2}- \frac{3}{h^3}\left( 2 y_2 + \frac{I_2-2K}{I_3} \right) \right] [a(x+\phi)+a(x)] \nonumber \right.\\
&\left.-\frac{1}{1+a(x+\phi)-a(x)}\right\}\d x \ ,
\end{align}
and $f_x^s$ refers to the force  in Eq.~\eqref{force}.  The  change in the phase angle between the two sheets is therefore proportional to the force which would be acting between them if they were prevented from having any relative motion. 
For small-amplitude waves,  we get $\Lambda^{-1}\approx 2\pi$,  so that near the fixed points, the phase behaves as  $\dot{\phi'}\sim \pm \delta^4\phi'^3$ and  we have $\phi'\sim t^{-1/2}$ near stable fixed points and $\phi'\sim (\tilde t-t)^{-1/2}$ near unstable points.

\begin{figure}[t]
\includegraphics[width=0.46\textwidth]{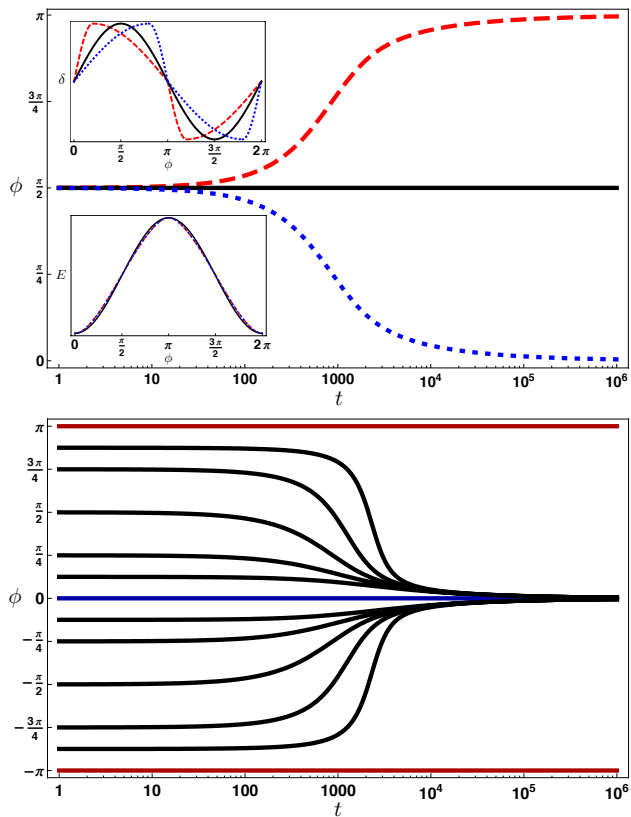}
\caption{(Color online). Top: Evolution of the phase angle, $\phi$,  from an initial value of $\pi/2$ for swimmers with three different waveforms [upper inset]: sinewave (black solid line), sinewaves shifted forward and backwards by $3\pi/10$ (blue dotted and red dashed lines respectively). As predicted by the theory, the perfect sinewave shows no phase evolution  while the skewed sine waves evolve to phase-locked positions (in-phase and opposite-phase respectively). The energy dissipated for the three waveforms is almost identical 
[lower inset]  (see text). 
Bottom: For all initial conditions,  the phase angle converge to a phase-locked state at the only stable fixed point, $\phi=0$ (same parameters as in top figure, dotted line).
}
\label{phi_energy}
\end{figure}

As discussed above, the geometry of the wave is the only factor determining the direction in which the relative position of the sheets evolve. For illustration, we now consider waves in the form of skewed sinusoids. We map  a sinewave  from the intervals $[0:\pi/2]$ and $[\pi/2:\pi]$ to the  intervals $[0: \pi/2+\alpha]$ and $ [\pi/2+\alpha:\pi]$, and  define $a(x+\pi)=-a(x)$ on the interval $[\pi: 2\pi]$. Shapes with $\alpha>0$ ($\alpha<0$) have a larger region where the wave amplitude increases (decreases) in the direction of the wave propagation (see  Fig.~\ref{phi_energy}, upper inset). If $\alpha>0$ ($\alpha<0$) then $A<0$ ($A>0$) and our analysis predicts that phase-locking will occur at the in-phase (opposite-phase) conformation. Experimental observations in Ref.~\cite{rikmenspoel} suggest the flagellar amplitude of bull spermatozoa is reasonably given by a linearly increasing sine wave; this yields front-back asymmetry corresponding to $A<0$. We now proceed to solve Eq.~\eqref{phi_dot} numerically.

The dynamics of phase-locking is illustrated in Fig.~\ref{phi_energy} (top), where we plot the evolution of the phase angle from an initial phase $\phi=\pi/2$ for three shapes of swimmers (upper inset): perfect sinewave ($\alpha=0$, black solid line), and two skewed sinewaves with $\alpha=3\pi/10$ (blue dotted line) and $\alpha=-3\pi/10$ (red dashed line). The perfect sinusoidal shape yields no evolution in time, as predicted by the analysis. In contrast, the skewed sinewaves evolve into the predicted phase-locked positions,  in-phase for $\alpha>0$ and opposite-phase for $\alpha<0$ (at the same rate because the waveforms are symmetric to each other about the vertical axis).  The energy dissipated between the sheets, Eq.~\ref{energy},   is displayed  in  Fig.~\ref{phi_energy} for the three shapes [lower inset].
The dissipation for the two skewed waveforms is identical -- as $\dot E$ is invariant under a reversal of swimming direction --,  is slightly different from the perfect sinusoid (mean square difference $< 0.1\%$), and 
all shapes display the predicted maximum at $\pi$ and minimum at $0$ and $2\pi$. 
In agreement with our analysis, the numerical results demonstrate therefore that swimmers with  increasing amplitude ($\alpha>0$) phase-lock into the most energetically favorable conformation, while those with decreasing amplitude ($\alpha<0$)  phase-lock into the least energetically favorable conformation.  The example with $\alpha=3\pi/10$ is further illustrated in  Fig.~\ref{phi_energy} (bottom) where we show the  evolution of the phase angle  from various initial phases. In  all cases, the configuration evolves into a phase-locked state at the only stable fixed point, $\phi=0$. 
Further computations (not shown)  show that increasing the asymmetry of the waveform, or its amplitude, decreases the time scale over which the systems evolves into a phase-locked state.

In summary, in this paper we have used a simplified model to   show that hydrodynamic forces alone can lead to the observed phase-locking between two swimming microorganisms if their waveforms are front-back asymmetric. The nature of the phase-locked state, and the rate at which the relative conformation of the  two swimmers evolve to it, is  dictated solely by the geometry of the waveforms. In particular,  an in-phase conformation may be obtained when the swimmers  have shapes with increasing amplitude front to back, as observed for some mammalian spermatozoa \cite{rikmenspoel}. Other front-back asymmetries, such as the presence of a head, would also further contribute to phase-locking \cite{yang}.

Discussions with T. Powers and S. Spagnolie  and funding by the NSF (grants CTS-0624830 and CBET-0746285) are gratefully acknowledged. 

\bibliography{phaselocking}

\end{document}